\documentclass[article,twocolumn,showpacs,preprintnumbers,amsmath,amssymb]{revtex4}
\usepackage{graphicx}
\usepackage{dcolumn}
\usepackage{bm}
\usepackage{epsfig}

\begin{document}

\hyphenpenalty=100000

\title[In-medium hadronization]{In-medium hadronization in the deconfined matter at RHIC and LHC}

%
\author{R. Bellwied$^1$, C. Markert$^2$ }

\vspace*{.5cm}

\affiliation{$^1$ Physics Department, Wayne State University,
666 West Hancock, Detroit, MI 48201, USA \\
$^2$ Physics Department, University of Texas at Austin,
Austin, TX 78702, USA}

\email{bellwied@physics.wayne.edu}

\vspace*{1cm}

\begin{abstract}
We study the mechanism and probability of in-medium hadronization in the deconfined medium produced in heavy-ion collisions at RHIC and LHC. We show the likelihood of color-neutral objects to be formed inside the partonic fireball and the probability of these states to escape the medium with reduced interaction strength and energy loss. We will suggest specific measurements that are sensitive to the early degrees of freedom and show predictions for these measurements at RHIC and the LHC.
\end{abstract}



\maketitle


\section{Introduction}

Hadron formation has been an intractable problem in high energy physics since it was realized that particles form most abundantly in a kinematic regime that can not be described by perturbative QCD. Thus, the treatment of hadronization is largely based on phenomenological models that apply first principles such as energy conservation and the uncertainty principle, but that also try to incorporate insights derived from specific measurements. The question of how and when hadrons form has recently garnered new interest in particular when a deconfined thermal medium, such as the strongly interacting quark gluon plasma generated in heavy collisions at RHIC, is produced initially. We postulate that, to first order, hadronic formation times in a non-thermal fragmentation process can be treated independently from the hadronization of the thermalized surrounding medium. There is no a-priori reason why a hadron formed in a process that does not couple strongly to the medium should form at different times in vacuum and in medium. This idea has to be compared to the standard modeling of partonic energy loss in the deconfined medium.

In all existing energy loss models, which attempt to describe the hadronic suppression factors measured in heavy ion collisions at RHIC, the scattered high momentum parton is treated as a massless particle which will fragment only outside of the medium \cite{glv,asw,amy}. This allows the theory to treat the energy loss of the parton inside the deconfined medium as purely partonic. In these models the only dependence of the transport coefficient will be on the parton flavor, i.e. gluons should lose more energy than quarks (due to the Casimir factor \cite{casimir}) and light quarks should lose more energy than heavy quarks (due to the dead-cone effect \cite{kharzeev}). Unfortunately, initial measurements of heavy quark energy loss through semi-leptonic decay channels have shown little evidence for the dead cone effect \cite{star-heavy,phenix-heavy}, and the gluon vs. quark energy loss imbalance is also not borne out in the data \cite{star-midpt}, although this measurement might be hindered by the fact that it is not (yet) possible to isolate pure gluon and quark jets at RHIC energies. Nevertheless the apparent agreement of the nuclear suppression factors of light mesons and leptons from heavy meson decay forced theorists to re-evaluate the balance between radiative and collisional energy loss \cite{djord}.  This leads to an explanation for the measured electron suppression as long as the vast majority of these electrons is due to D-meson and not B-meson decay, which only holds true at sufficiently small transverse momentum. An alternative explanation has been offered by Adil and Vitev who calculated the probability that a D-meson forms early in the medium and then dissociates through an enhanced meson-parton interaction probability \cite{adil, vitev}. They concluded that the process of formation and dissociation can occur several times during the partonic lifetime of the medium and that this process would enhance the suppression factor of all final state heavy mesons. The underlying theory of hadron formation inside the deconfined medium prior to reaching the phase boundary is also the focus of our study. We will show that the boost, which is considered the primary cause for high momentum partons to fragment outside of the medium is partially offset by energy conservation requirements in the fragmentation process. The resulting production time of a color-neutral object can be well within the extended lifetime of the deconfined medium at RHIC and LHC energies, which means that not all (pre-)hadrons form at the critical temperature (hadronization surface), but some of them, in particular the ones from non-thermal fragmentation processes can form well ahead of bulk matter hadronization. 

In the following sections we will discuss previous modeling of early formation times in comparison to our own calculation of p$_{T}$ dependent formation times of pre-hadrons. We will then introduce the relevant difference between colored quasi-particles and color-neutral pre-hadrons, as well as the concept of color transparency. The relation to old cluster formation theories in this context will also be discussed.  Finally we will present a list of possible signatures of in-medium hadronization including nuclear suppression factors and other experimentally verifiable effects.

\section{Early formation time models}

The initial formulation of the hadron formation time in a high momentum fragmentation process was based on Heisenberg's uncertainty principle. Due to the boost attained by particles formed in collisions at relativistic energies, it was concluded that the minimum formation time depends linearly on the $\gamma$-factor and is inversely proportional to the mass of the particle. This led to the concept of the inside-outside cascade in the parton fragmentation process, i.e. the higher the momentum of the particle the further 'out' it is formed. This basic concept was first augmented by Bialas and Gyulassy, who suggested that an outside-inside component to this process has to be taken into account if the fragmentation occurs in medium \cite{bialas}. The main reasons were worked out in detail later on by Kopeliovich and collaborators \cite{kope1}. The key argument is that, due to energy conservation, a parton fragmenting inside a medium into hadrons with fractional momenta z has to generate high z particles first, simply because in-medium effects such as bremsstrahlung, will lead to partonic energy loss. In other words, when the initial parton loses too much energy the high z hadrons can not be formed anymore. Brodsky and Mueller therefore suggested to distinguish between the production time ($\tau_{p}$) of the color-neutral object of a fixed fractional momentum  which is driven by energy conservation, and the subsequent formation time ($\tau_{f}$) of the final hadronic wave function, which is driven by the boost \cite{brodsky}.  Fig.1 (Scheme A) shows a time evolution of the hadronization process based on these principles as proposed by Accardi \cite{accardi}.

\begin{figure}[!t]
\includegraphics[width=3.in]{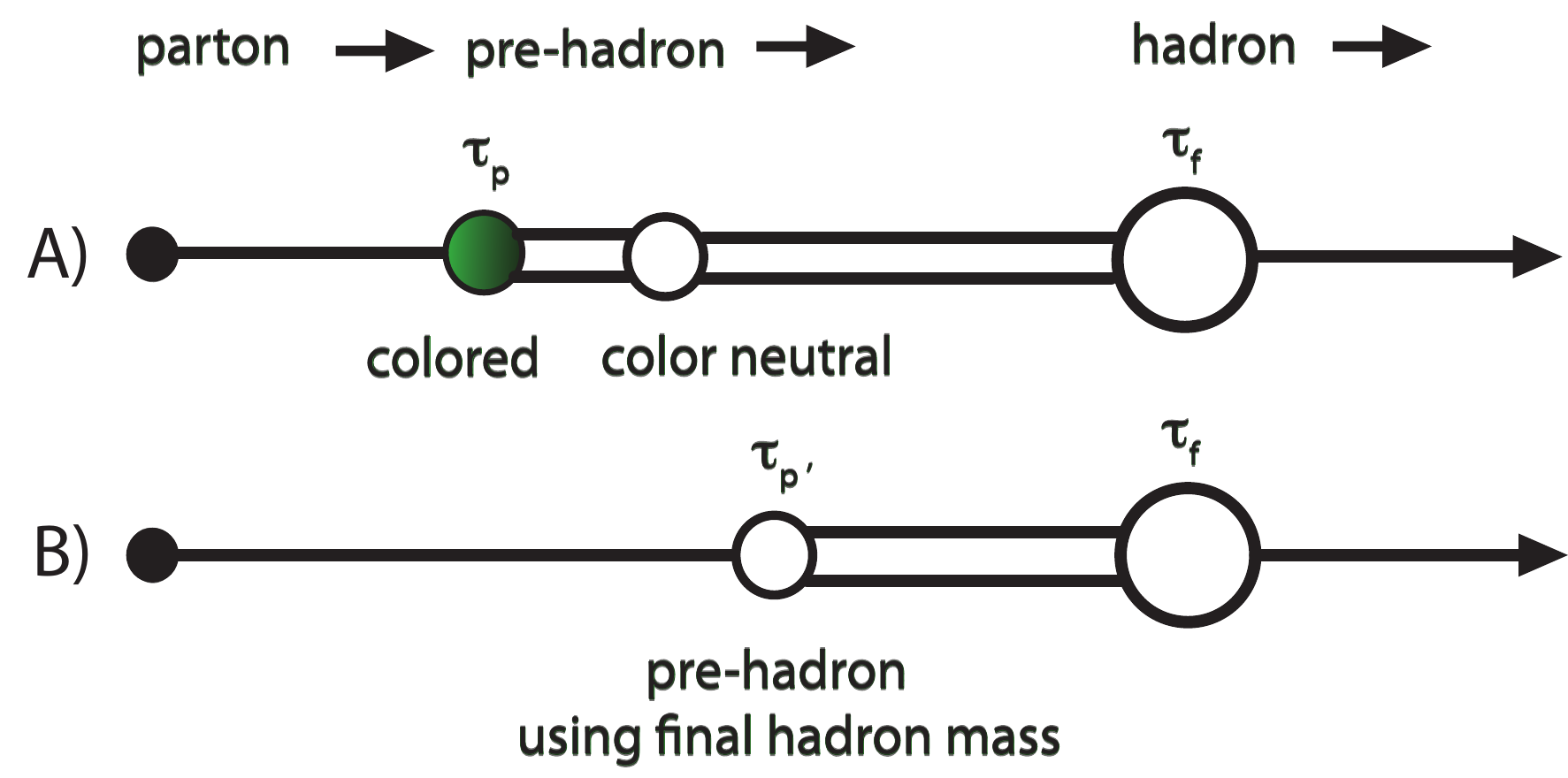}
\caption{\noindent Schematic evolution of the hadronization process from an initial parton to the final hadronic state \cite{accardi}. Scheme B displays the variation proposed in our approach \cite{mbv}.}
\label{form1D}
\end{figure}

A detailed study of the difference between $\tau_{p}$ and $\tau_{f}$ in the framework of string models, such as PYTHIA and JETSET, was presented by Gallmeister and Falter \cite{gallmeister}. In such models the production time depends only slightly on the final hadron mass and thus reflects the production of a massless color-neutral object which does not yet carry any hadronic features. The question of when color-neutrality is obtained in the pre-hadronic phase can not be calculated and thus one can assume that at the time of production $\tau_{p}$ a color neutral pre-hadron is formed, which is slightly different from Accardi's scheme. The other boundary is the formation time, when all hadrons have attained the full wave function and hadronic properties, which is dominated by the gamma factor.

We have proposed a comprehensive approach to the hadron formation \cite{mbv}, which models the interplay between the boost and energy conservation by calculating an intermediate complex formation time ($\tau_{p}$' ) between  $\tau_{p}$ and $\tau_{f}$ as is shown in Fig.1 (Scheme B). This time can be calculated based on light cone variables which depend on the final hadron mass:

\vspace{-0.3cm}
\begin{eqnarray}
\Delta y^+ &= & \; \frac{zp^+}{m_h}   \times  2 \left[ m_h +
\frac{ {\bf k}^2}{(1-z)m_h} - \frac{zm_q^2}{m_h}
                              \right]^{-1} \; \quad
\label{tfrag}
\end{eqnarray}

Here $p^+$ is the large lightcone momentum of the parent parton and $|{\bf k}| \sim \Lambda_{QCD}
\sim 200$~MeV/c  is the deviation from collinearity. The  lightcone  $\Delta y^+ =  \tau_{\rm form} +
l_{\rm form}$  is conjugate to the non-conserved lightcone momentum component $\Delta p^- = (p^-)_{final} - (p^-)_{initial} $ of the evolving system. The formation time then reads:

\vspace{-0.3cm}
\begin{eqnarray}
&& \tau_{\rm p'} = \frac{ \Delta y^+}{ 1+\beta_q} \;, \quad
\beta_q = \frac{p_q}{E_q} \;
\label{tform}
\end{eqnarray}

Both Kopeliovich and Accardi have pointed out that such a formation time does not have to describe the final state on-shell hadron but rather describes the color-neutral object which has not developed the full hadronic wave function yet. That means the width and thus the interaction probability of this initial state is small and will build up only over time, i.e. potentially in the vacuum outside the medium.
We postulate that our calculation yields the earliest moment at which the color-neutral pre-hadron attains its spectral function and will exhibit hadronic properties such as mass, width and specific decay branching ratios. Further details of the derivation can be found in \cite{mbv}.

Regarding the interaction probability with the surrounding colored medium these pre-hadrons shall experience considerably less radiative energy loss since the condition for induced gluon emission is not met by the traversing object. We also argue that the collisional energy loss will be minimized due to color transparency in medium for small sized color-neutral pre-hadrons. 
In other words, the decreased partonic attenuation probability of these states in the medium will likely lead to less energy loss and thus a smaller nuclear suppression factor (R$_{AA}$) at high p$_{T}$. Our original paper emphasized the effect of early formation on hadronic resonances \cite{mbv}. Here we show that non-thermal fragmentation to high momentum hadronic ground states could lead to a mass dependence of the suppression factors measured at RHIC and LHC, which can be explained by assuming early formation of color-neutral pre-hadrons in the partonic phase.

Accardi \cite{accardi} and Cassing/Bratkovskaya \cite{cassing} have applied similar ideas in different ways. Accardi has used the color-neutral pre-hadron concept to describe particle identified data measured in eA collisions at HERMES \cite{hermes2}, Cassing and Bratkovskaya have used a dynamical quasi-particle model (DQPM) coupled with parton-hadron string dynamics (PHSD) to describe measured data. Here a dressed colored state evolves in the medium rather than a color-neutral object. In the following section we will show that this slight distinction of the state prior to hadronization might lead to significant differences in the resulting hadron spectra.

\section{The concept of clusters, quasi-particles, pre-hadrons and color transparency}

The main question regarding early formation of pre-hadrons is whether such a state can indeed be considered hadronic at any particular time during its evolution. Following the formation time of a color-neutral state during parton fragmentation according to Kopeliovich et al. \cite{kope1} yields a time that is about half of the final hadron formation time. This pre-hadronic state might only turn color-neutral some time after its formation (see Fig.1), but even then the state has to attain its width before it can be considered an actual hadron. As soon as the state has obtained a certain width, it can interact strongly on a timescale inverse to the width. This could be a collisional width, in which case the particle elastically or inelastically scatters so that its energy is uncertain by the interaction rate.  The probability of interaction will rise with the width and thus one needs to balance the generation of the hadronic wave function with the requirement for a reduced interaction cross section. Alternately the particle could decay into other states before attaining its final width as was postulated for the case of partially chirally restored hadronic pre-resonances \cite{mbv}.

The interaction cross section between a color-neutral state and a colored medium is very small and potentially close to zero if color transparency can be assumed. The concept of color transparency was introduced by Brodsky et al. \cite{brodsky, brodsky-transp,jain} and, in the strictest sense, only applies to hadrons produced directly in higher twist parton-parton subprcesses. Only then can a color neutral configuration be considered point-like, which leads to the minimal cross section with the surrounding colored medium.  Sickles and Brodsky \cite{sickles} have applied color transparency to direct proton formation at RHIC energies and found that it can explain the unusually large baryon to meson ratios at intermediate p$_{T}$ in RHIC heavy ion collisions \cite{phenix-midpt, star-midpt}.

We propose that a composite pre-hadronic state which is formed early, but not directly, will also exhibit some form of color transparency, because the hadronic wave function is not fully developed and thus the state is 'small'. This is partially confirmed through measurements in cold nuclear matter \cite{accardi,hermes2}. Jain et al \cite{jain} have pointed out that in the time evolution of the color-neutral object the interaction probability can be further reduced through destructive interference among the different hadronic modes in the wave packet. Ultimately the continuous 'dressing up' of the state over the period of the lifetime of the deconfined medium will lead to enhanced partonic attenuation, but only near the phase boundary after the color-neutral object has almost fully traversed the medium. This is not only due to initial color transparency but also the medium expansion which proceeds with a smaller collective velocity than the velocity of the high momentum fragment. This decoupling should lead to a relatively modest attenuation close to T$_{c}$ which will affect the ground state hadronic configurations only in a negligible way. An exponentially growing attenuation of the pre-hadronic state would be relevant only in the hadronic corona where medium modification effects are small \cite{kope2}. One exception is the enhanced partonic attenuation of wide hadronic pre-resonances which might lead to partial chiral restoration.

It needs to be noted that alternative approaches to the hadronization mechanism mostly differ in their evolution of the degrees of freedom with respect to reaching color neutrality. In the model by Cassing and Bratkovskaya \cite{cassing}, which is based on early quasi-particle formation, the colored state attains a dynamic color field, i.e. a dynamic mass, until hadronization. These states are thus dressed gluons and quarks i.e. no composite objects. The colored quasi-particle then recombines near T$_{c}$ into a very heavy color-neutral quasi-particle (color neutrality is not only achieved in 2 and 3-quark configurations) which subsequently decays into several final hadronic states of lesser mass. Although this picture includes the recombination of quasi-particles it does not violate entropy to the degree of constituent quark recombination, i.e. the recombination of thermal quarks into hadronic states of correct mass. 

A similar approach to hadronization is applied in cluster models  which were originally developed in the 70's in order to describe not only particle production but also the correlations betwen particles measured in elementary collisions \cite{foa}. The cluster model event generator, HERWIG \cite{herwig}, was more
successful in describing correlation features than its fragmentation counterpart, PYTHIA \cite{pythia}. In HERWIG a cluster of very large mass (colored or color-neutral) is formed and then decays into final state hadrons. No particular production mechanism for the clusters has been suggested, but the masses that are in best agreement with the correlations measured are on the order of 3 GeV/c$^{2}$. In the past, several authors (e.g. \cite{shuryak} and references therein) have indicated that these clusters could be instantons or sphalerons.  These massive partonic clusters could be, upon transitioning into the hadronic phase, identical in their quantum properties to Hagedorn states, i.e. very large mass hadronic resonances which decay strongly into lower mass resonances or hadronic ground states (baryons and mesons) \cite{noronha}. 

Shuryak and others have postulated that these states could be formed (or survive depending on the initial state) above the critical temperature, up to a temperature of 1.2 T$_{c}$ \cite{shuryak2}. From our point of view though, if any composite states are supposed to be the relevant degrees of freedom for the strong coupling phase in the sQGP, they need to exist above 1.5 T$_{c}$ in order to explain the observed strong anisotropic flow. In that sense pre-hadronic clusters that form in the partonic phase are more likely to explain the elliptic flow than surviving Hagedorn states above T$_{c}$. The latest lattice results of the deconfinement strength based on renormalized Polyakov loop calculations are shown in Fig.2 \cite{petreczky}. Full deconfinement is only achieved above 2.5 T$_{c}$ and the rather smooth evolution of L$_{ren}$ can be viewed as a measure of free quark suppression in an extended transition (crossover) region. In other words, there is a mixed phase of degrees of freedom where the free quark suppression is smoothly taken up by the formation of pre-hadrons (either color-neutral objects or colored quasi-particles) as indicated in Fig.2.

\begin{figure}[!h]
\includegraphics[width=3.in]{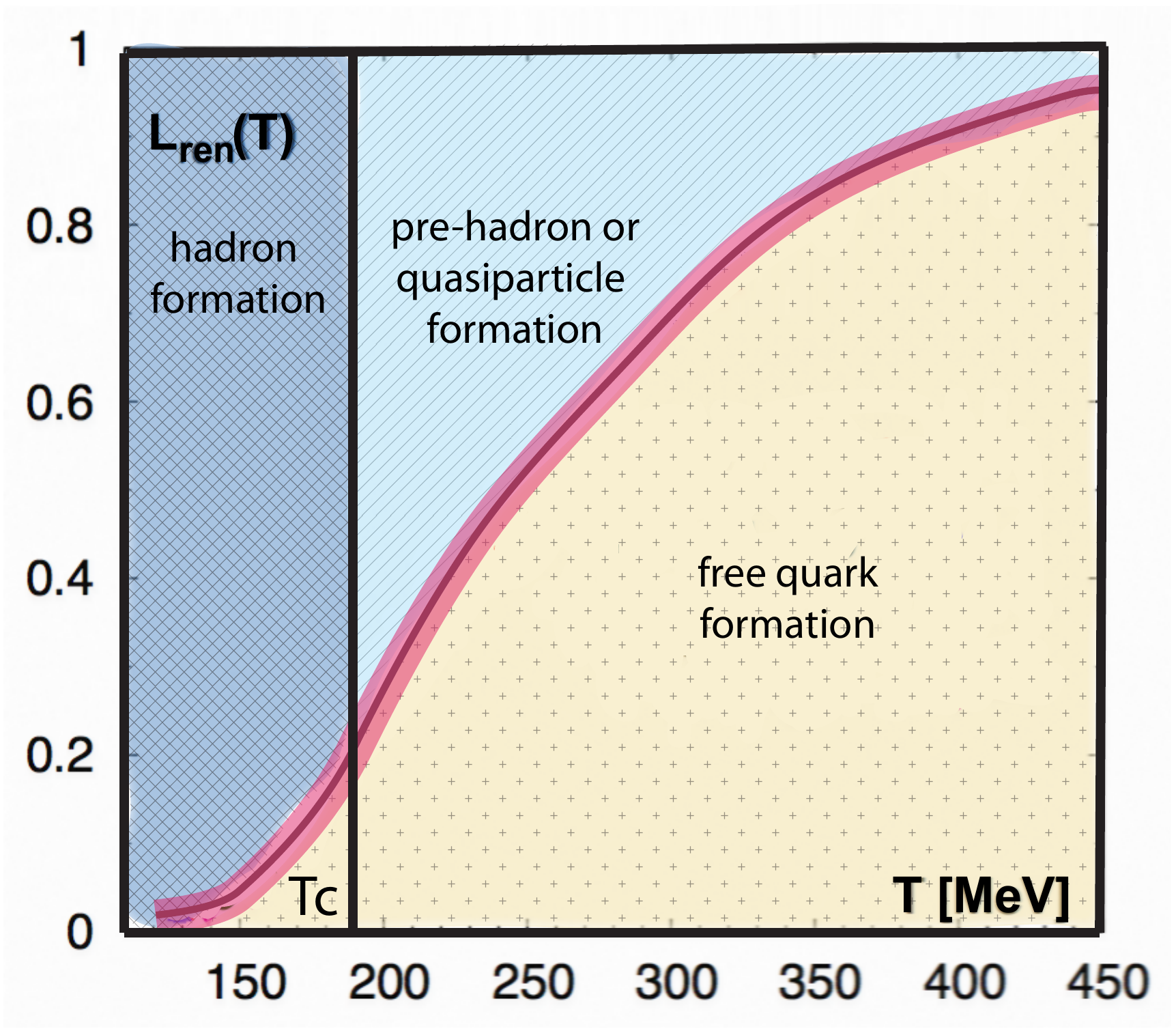}
\caption{\noindent The curve represents lattice calculations for the level of deconfinement based on renormalized Polyakov loops \cite{petreczky}. The shaded areas represent our understanding of the relevant degrees of freedom as a function of the temperature near T$_{c}$.}
\label{form1D}
\end{figure}

A distinguishing characteristic between colored quasi-particles and color-neutral pre-hadrons is that in quasi-particle models the correspondence of a particular partonic state to a final hadron is not defined until final hadronization, which means these models should not lead to some of the tangible experimental differences (e.g. nuclear suppression factors) for different final state hadrons, which we propose in the next section of this paper. Regarding the relative likelihood of these two competing mechanisms it is safe to say that although the high momentum fragmentation process leads naturally to color neutral pre-hadron formation, the bulk matter production is likely dominated by colored objects which either recombine or hadronize thermally. Therefore most of our experimental signals for pre-hadron formation will be based on high momentum probes.

\section{Signatures for in-medium hadronization}

The concept of a reduced cross section of a pre-hadron in a hadronic medium was applied to deep inelastic scattering on nuclei (eA) data from DESY, in particular the HERMES experiment \cite{hermes}. Two experimental signatures were proposed: reduced p$_{T}$-broadening and reduced nuclear attenuation of identified particles in cold nuclear matter. Qualitatively, the measured broadening and the attenuation are smaller than predicted for the propagation of a bare parton and thus are in agreement with pre-hadron formation. Accardi et al. \cite{accardi} as well as Falter et al. \cite{fgm} have used these data to evaluate the formation times of the color neutral states, and their results are
within the uncertainties comparable to our estimates \cite{mbv}. The main difference of the HERMES measurements to propagation in QGP matter would not be in the formation time, but rather in the interaction probability of the pre-hadron. If color transparency is the dominating factor than both p$_{T}$-broadening and attenuation (or better suppression) at high momentum should be as reduced at RHIC than at DESY when compared to models that assume purely partonic propagation through the cold nuclear or hot dense partonic medium. Although the nuclear suppression factor increases fifteen fold from HERMES to STAR/PHENIX, the relative difference between parton or pre-hadron propagation should be roughly constant. In \cite{mbv} we have shown that the production times for all high $p_{T}$ hadrons from fragmentation (p$_{T}$ $>$ 2 GeV/c), except the pions,  are well within the lifetime of the deconfined medium at RHIC and LHC up to a certain maximum transverse momentum. As an example Fig.3a shows that at RHIC Kaons up to 10 GeV/c and protons up to 20 GeV/c are formed in the medium. The medium lifetime was derived analytically, but it is worthwhile noting that the calculation is in good agreement with empirical estimates based on HBT and resonance yield measurements \cite{hbt, markert1}. For an estimate of the effect at higher energies we note that the QGP lifetime is expected to be roughly twice as long at the LHC \cite{mbv}.

The early formed pre-hadrons should exhibit a very small interaction cross section, depending on their formation time, which in our model depends on their final hadron mass. Thus we propose three experimentally verifiable effects:

1.) The nuclear suppression factors (R$_{AA}$) should reduce according to the production time of the pre-hadron. The earlier a color neutral object is formed the less attenuation it should exhibit. In the case of pions, kaons and protons this should lead to different p$_{T}$-dependent  R$_{AA}$ factors even when only the fragmentation process is considered. 

\begin{figure}[!h]
\includegraphics[width=3.in]{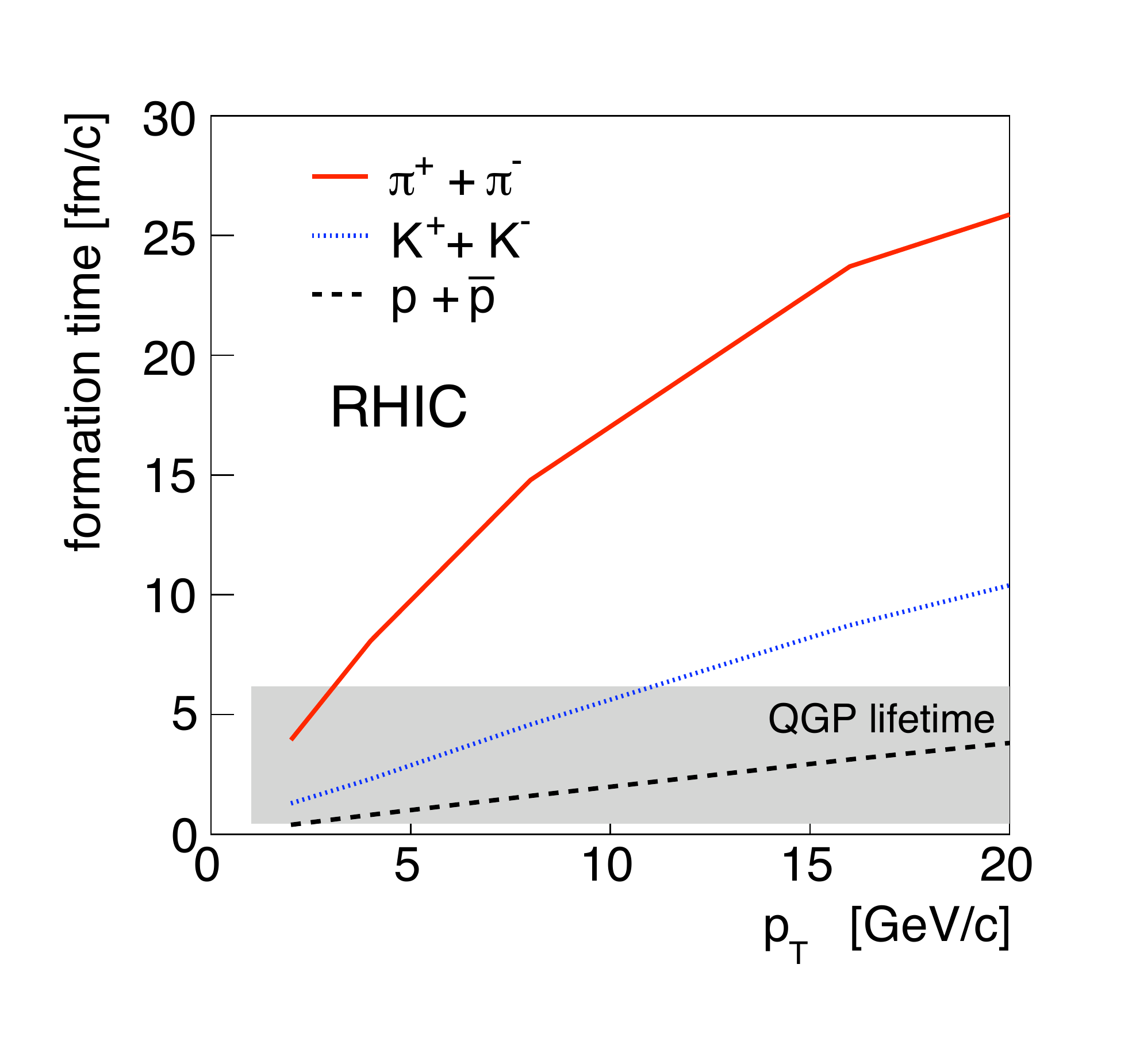}
\includegraphics[width=3.in]{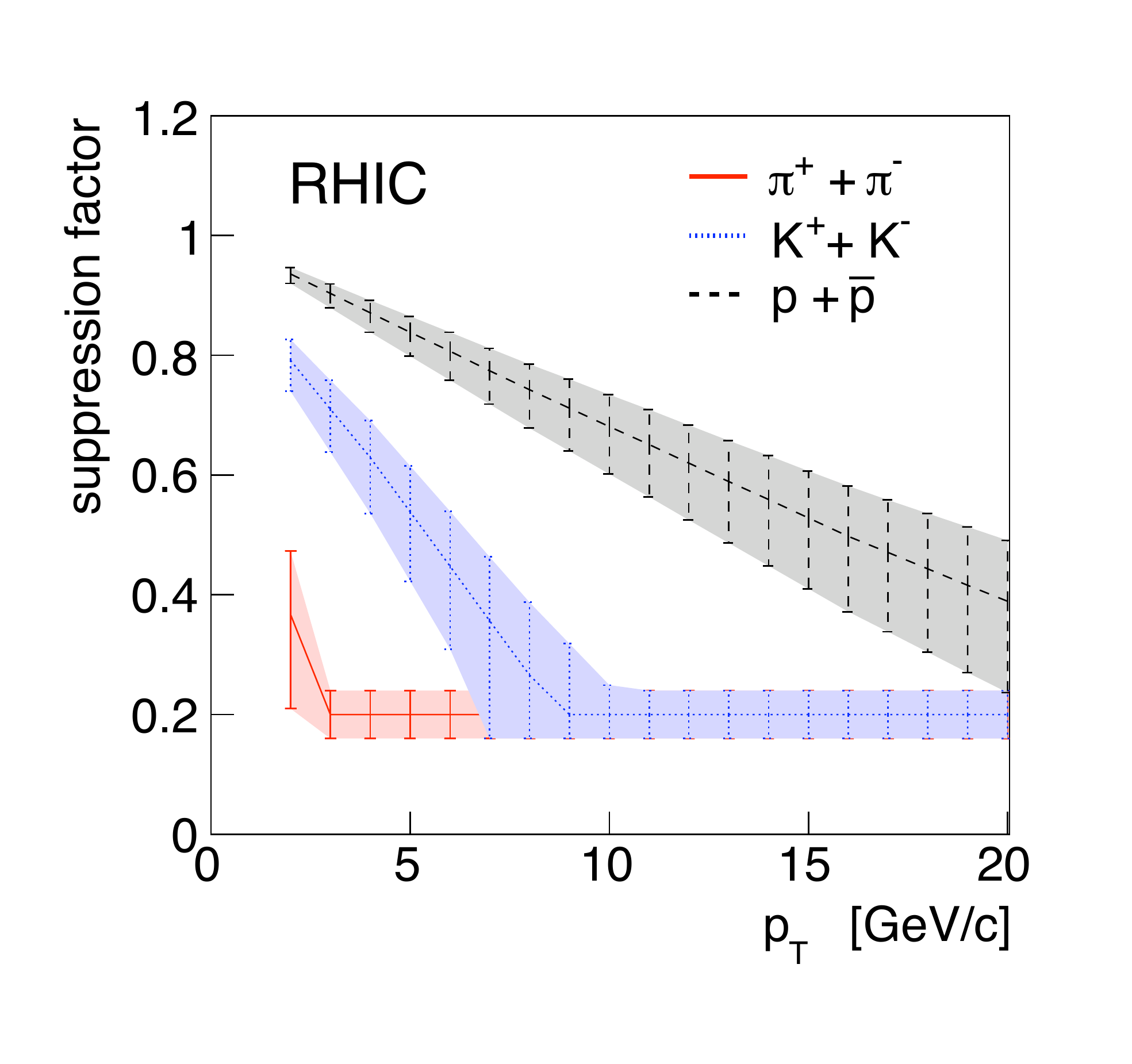}
\caption{\noindent a.) Momentum dependence of formation times of pions, kaons and protons at RHIC energies in comparison to QGP lifetime according to \cite{mbv}. b.) predicted nuclear suppression factors (R$_{AA}$) for pions, kaons and protons based on their formation times in the medium formed in central 200 GeV/n Au+Au collision at RHIC.}
\label{form3D}
\end{figure}

Fig.3b shows the resulting suppression factors under the assumption that a hadron formed very early in the plasma phase shows no suppression due to color transparency and a hadron formed after the plasma phase (i.e. in vacuum) shows maximum suppression due to partonic energy loss. The uncertainties shown are based on a QGP lifetime variation of 5$\pm$1~fm/c. The minimum R$_{AA}$ was set to 0.20$\pm$0.04 for pions based on an average of measurements by PHENIX \cite{phenix-pions} and STAR \cite{star-pions}.

The calculation can also be applied to LHC energies, which is shown in Fig.4. Here we assumed a QGP lifetime of 14$\pm$2 fm/c and a maximum pion suppression factor equal to the one at RHIC. 

\begin{figure}[!h]
\includegraphics[width=3.in]{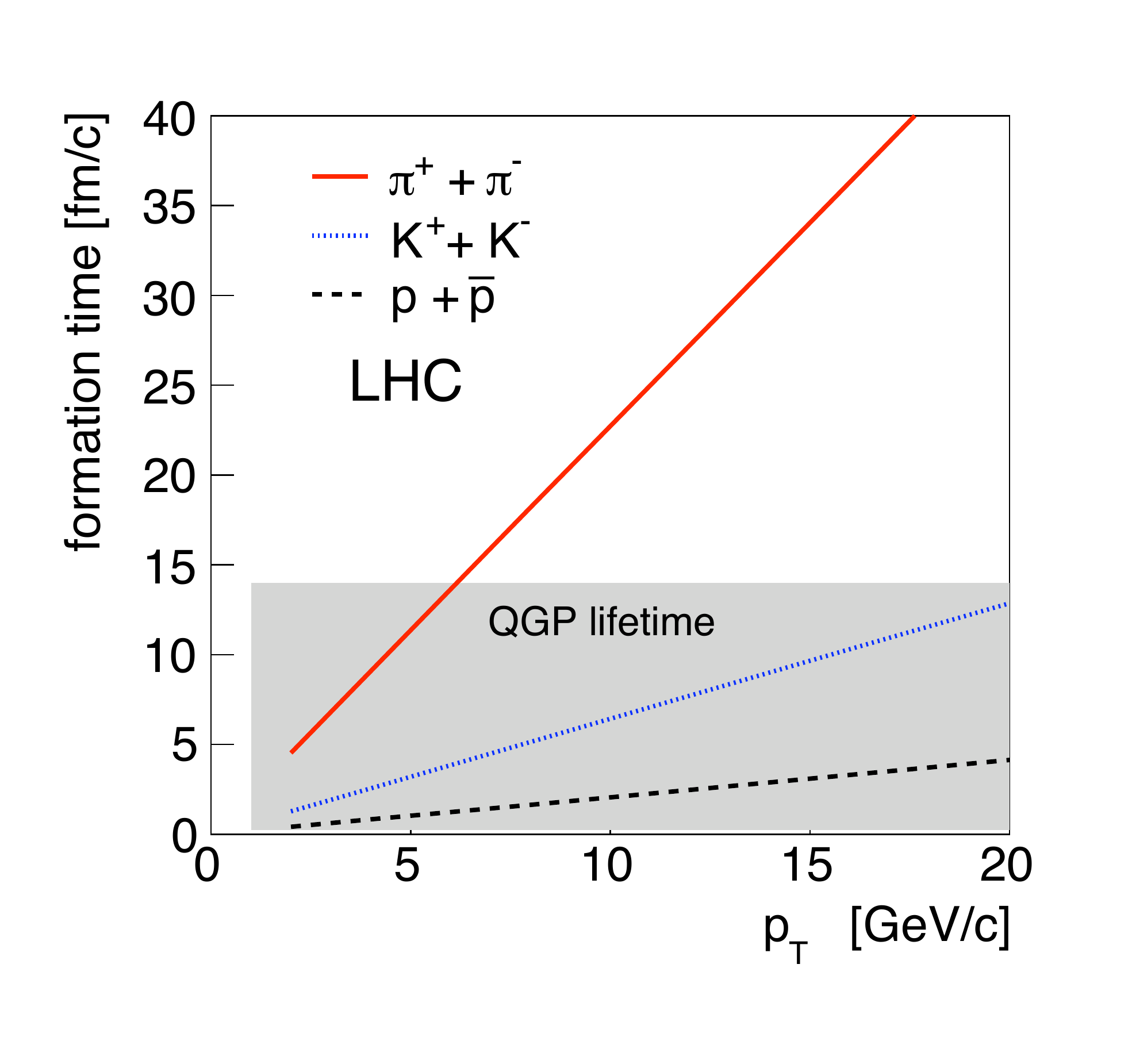}
\includegraphics[width=3.in]{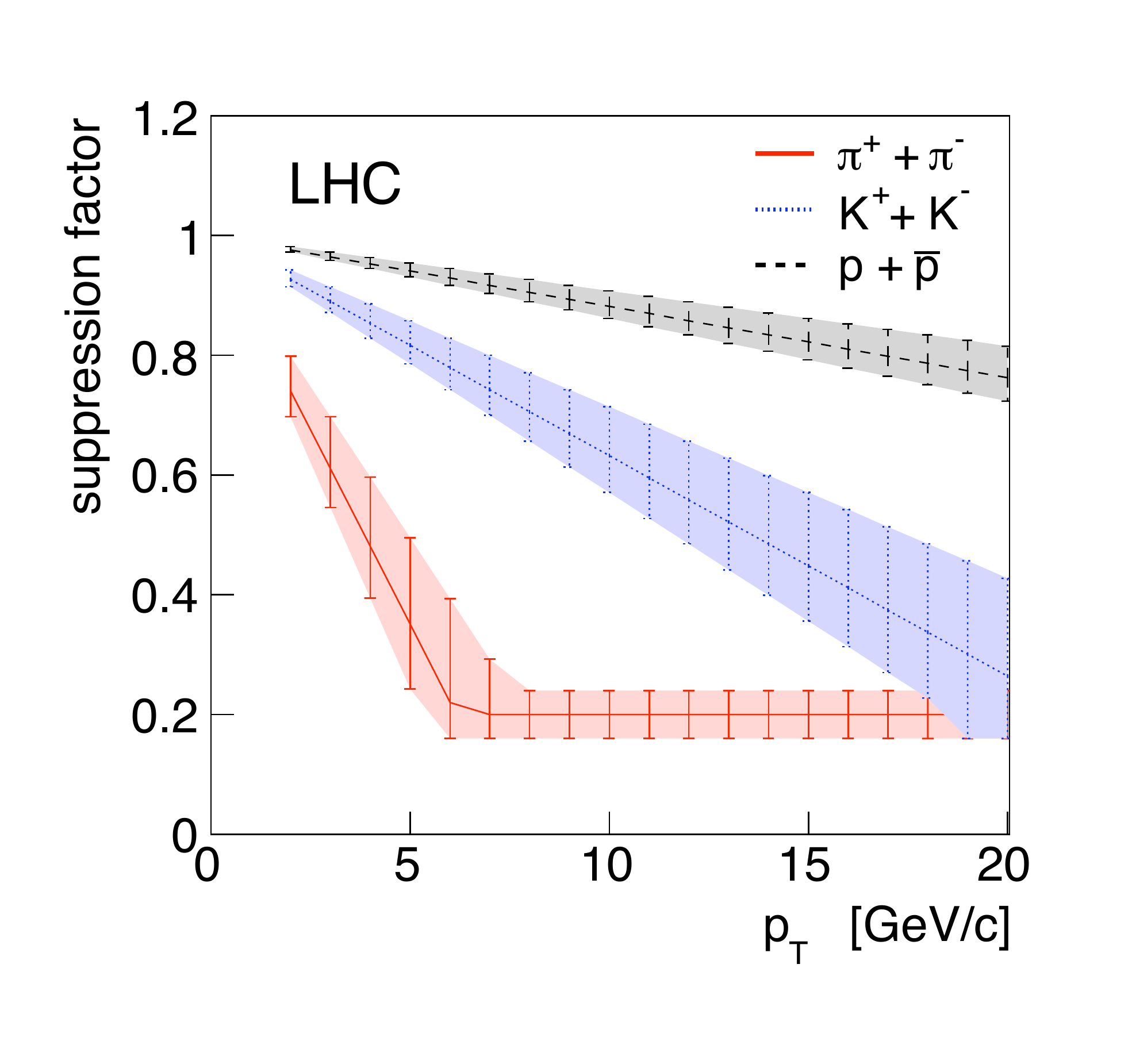}
\caption{\noindent a.) Momentum dependence of formation times of pions, kaons and protons at LHC energies in comparison to QGP lifetime according to \cite{mbv}. b.) predicted nuclear suppression factors (R$_{AA}$) for pions, kaons and protons based on their formation times in the medium formed in central 5.5 TeV/n Pb+Pb collision at LHC.}
\label{form3D}
\end{figure}

Measuring the maximum effect shown here is predicated on the assumption that particle production is dominated by fragmentation down to low momentum and that a color neutral object is color transparent throughout the lifetime of the partonic phase. In that sense our reduced suppression factors are strict upper limits. In addition, we know from data that it is likely that fragmentation is not the only hadronization mechanism in particular at lower p$_{T}$ where recombination and statistical hadronization are dominant. Thus the results shown here should not be taken too literally below p$_{T}$ = 6 GeV/c. But in the fragmentation region (p$_{T}$ $>$ 6 GeV/c) protons should be suppressed less than pions and kaons (R$_{AA}$(p) $>$ R$_{AA}$(K) $>$ R$_{AA}$($\pi$)) until their momentum exceeds the limit for in-medium hadronization. 
In this regime light quark resonances ($\rho$, $\eta$, $\omega$) might still show similar suppression than pions simply because the pre-hadron states can be linear combinations of light quark condensate states and thus a pion can only be distinguished from an $\eta$ or a $\rho$ after the final hadronic wave function has developed. But strange quark resonances (K$^{*}$, $\phi$, $\Lambda^{*}$) should exhibit less suppression than light quark resonances for example. Alternately for all pre-resonances the off-shell generation leads to a significantly reduced lifetime and broadening of the state which means the resonances are likely decaying inside the medium and interact strongly in the medium. Thus their energy loss should be comparable to the one obtained for colored objects (partonic energy loss) as long as they are reconstructed with their ground state mass.

Similar suppression factors for light and heavy mesons can be explained by an enhanced dissociation probability for specific pre-hadrons in the partonic phase. In the approach by Adil and Vitev \cite{adil, vitev} the heavy meson does not exhibit color transparency or even a reduced interaction cross section in the deconfined medium, but rather dissociates and regenerates several times throughout the lifetime of the  QGP. Since the heavy pre-meson production is even faster than the light pre-baryon production, the temperature and density of the QGP is such that dissociation is more likely for these states. In our calculation of the light quark states we ignore dissociation and assume maximum color transparency.
In this context Greiner and collaborators  \cite{ggc,zxu} have speculated that final state interactions of leading light pre-hadrons with the partonic medium could also account for a large fraction of the measured energy loss in heavy ion collisions, as long as no color transparency is assumed and the pre-hadron formation time is fixed to less than 0.8 fm/c, which is unrealistically small for most produced states according to our calculations.

Theories based on only parton propagation in the medium and hadronization in vacuum can, under certain conditions, also yield differences in the nuclear suppression factors for light hadrons. Liu and Fries \cite{lf} have suggested that the quenched parton will undergo flavor conversions due to inelastic interactions in the dense partonic medium. They showed that this could affect the kaon suppression compared to the pion suppression at high p$_{T}$.  Sapeta and Wiedemann \cite{sw} have suggested that large gluon splitting as the mechanism for partonic energy loss could lead to hadronic yields which scale with the mass of the hadron. Their fragmentation modification factor and its dependence on the hadron mass was an ad-hoc parametrization, though, which still requires better justification. 

2.) High momentum protons should exhibit less p$_{T}$-broadening than pions because the pre-proton is formed inside the medium, whereas the pre-pion is formed in vacuum and its broadening is thus dominated by the strong interaction between the initial quark and the colored medium. Since energy loss and p$_{T}$-broadening are closely connected \cite{baier}, any reduction in the nuclear suppression factor through early color-neutrality in the partonic phase should also lead to smaller p$_{T}$-broadening of the final state hadron. This effect has been studied in detail in cold nuclear matter based on HERMES results \cite{hermes} which led to the first experimental verification of early hadronic formation times in medium \cite{accardi}. The data shows that color-neutrality is achieved only near the surface of the surrounding cold nuclear matter in these deep inelastic scattering reactions, which is not surprising since the kinematic range of the produced hadrons and the lack of a co-moving fireball prohibits a longer co-existence time. Still the fractional momentum dependence of the p$_{T}$-broadening effect in these measurements reveals that the highest z particles are produced well within the surrounding matter.

Similar measurement will be considerably more difficult in partonic matter for several reasons. First, the hadron-parton interaction probability is much less studied than the hadron-hadron interaction. Furthermore
the transverse hadron momentum p$_{T}$ is defined relative to the direction of the jet. The p$_{T}$-broadening $\Delta$$<$p$_{T}^{2}$$>$$^{h}_{AA}$ is defined as the difference of the average squared transverse momentum of a detected hadron h produced in an AA interaction compared to that in an elementary proton proton interaction:

\vspace{-0.3cm}
\begin{eqnarray}
\Delta<p_{T}^{2}>^{h}_{AA} = <p_{T}^{2}>^{h}_{AA} - <p_{T}^{2}>^{h}_{pp}
\end{eqnarray}

The unambiguous interpretation of $\Delta$$<$p$_{T}^{2}$$>$$^{h}_{AA}$ will be difficult due to effects such as
the gluon radiation of the colored parton prior to pre-hadron formation, soft multiple interactions of the
color neutral state, and late interactions of the final hadrons with the co-moving hadronic medium. Furthermore
the definition of the jet cone and the jet axis are very complex in the large background generated in heavy ion collisions. Still, the magnitude of the expected effect should directly correlate with the postulated change in nuclear suppression shown in Figs.3b/4b.

3.) In-medium formed off-shell resonances should exhibit enhanced width broadening and potentially mass shifts. These possible signatures of chiral restoration were discussed in detail in our previous publication \cite{mbv}. Here we just note that depending on the production mechanism and the formation time the pre-resonance could be at a lower or higher mass than the on-shell final state. These high momentum, early formed pre-resonances from fragmentation, which also observe some level of color transparency according to our calculation, will not exhibit large suppression and their reconstruction from the measured decay products should reveal medium modified properties.

In summary, both RHIC detectors and ALICE at the LHC have PID capabilities significantly beyond 6 GeV/c, and it is possible that the initially considered featureless fragmentation region, which showed similar suppression for pions and heavy mesons based on their semi-leptonic decay products, might reveal suppression differences even in the light quark sector upon more detailed studies. Preliminary particle identified results from RHIC,with rather large experimental error bars, indicate first deviations from a common partonic energy loss picture \cite{xu,ruan,kijima}.

Pre-hadron formation might also be the cause for the apparent discrepancy between elliptic flow and nuclear suppression. Overall models that assume zero or even a small finite mean free path yield the correct v$_{2}$, but overestimate the suppression by orders of magnitude. In a pre-hadron formation picture the v$_{2}$ would still grow very fast in the pure partonic phase, according to hydrodynamics, and probably be largely developed at the time of pre-hadron formation. The nuclear suppression factor, which develops throughout the phase above the critical temperature, would be reduced by the effect of the traversing color transparent pre-hadron.

\section{Summary}

We are suggesting that in-medium hadron formation above the critical temperature
could be the reason for several experimental puzzles measured at RHIC.
The proposed formation of color neutral pre-hadrons early on in the deconfined phase will not only explain particle dependent suppression factors, but also inconsistencies
in describing anisotropic flow and partonic energy loss within the same model.
Furthermore it will allow us to gain a deeper understanding of the hadronization process which
seems to be governed by the dynamics of fragmentation but also resembles striking features
of recombination, in particular in its scaling properties. We propose a color-neutral
pre-hadron as the relevant hadronization degree of freedom in a sizable region above the
critical temperature (T = 1-1.5 T$_{c}$) where lattice QCD calculations,
using renormalized Polyakov loops as a signature for the level of deconfinement, show a remarkable 
and as of yet unexplained suppression of color degrees of freedom (smooth and rather shallow crossover rather than steep transition from hadronic phase). We suggest that these missing free quarks can be found in color-neutral pre-hadronic clusters, which have a reduced interaction probability in the deconfined medium due to color transparency. In order to experimentally verify our theory we are proposing detailed measurements of particle specific nuclear suppression factors, the  p$_{T}$-broadening of identified particles in jets, and the properties of high momentum resonances from jets (yield, mass, width, lifetime) at RHIC and the LHC.

\vspace*{0.8cm}

\begin{acknowledgments}
We would like to thank Ivan Vitev for his calculations and his scientific input. We thank Wolfgang Cassing, Elena Bratkovskaya, Andre Peshier, Boris Kopeliovich, Stan Brodsky, Alberto Accardi, and Jorge Noronha for enlightening discussions on quasi-particles and pre-hadronic states.
This work was supported by U.S. Department of Energy Office of Science under contract numbers
DE-FG02-92ER40713, DE-FG03-94ER40845 and DE-SC0003892.
\end{acknowledgments}

\end{document}